\documentclass[twocolumn]{aastex61}
\usepackage{color}

\newcommand{\ie}{i.e.,\ }
\newcommand{\eg}{e.g.,\ }
\newcommand{\etal}{et~al.\ }
\newcommand{\kms}{km~s$^{-1}$}
\newcommand{\magsec}{mag arcsec$^{-2}$}

\def\muv{$\mu_V$}
\def\mub{$\mu_B$}

\def\bmv{$B-V$}
\def\re{$r_{\rm e}$}
\def\Msun{M$_\sun$\ }
\def\Lsun{L$_\sun$\ }
\def\magsec{mag arcsec$^{-2}$}
\def\brackmueb{$\langle \mu_B \rangle_e$}
\def\Leo~Group{{\color{red} REF}}

\begin{document}

\title{BST1047+1156: An extremely diffuse and gas-rich object in the Leo~I Group}
\shorttitle{BST1047+1156}
\shortauthors{Mihos \etal}

\author{J. Christopher Mihos}
\affiliation{Department of Astronomy, Case Western Reserve University, USA}

\author{Christopher T. Carr}
\affiliation{Department of Astronomy, Case Western Reserve University, USA}

\author{Aaron E. Watkins}
\affiliation{Astronomy Research Unit, University of Oulu, Finland}

\author{Tom Oosterloo}
\affiliation{Netherlands Institute for Radio Astronomy (ASTRON), The Netherlands}
\affiliation{Kapteyn Astronomical Institute, University of Groningen, The Netherlands}

\author{Paul Harding}
\affiliation{Department of Astronomy, Case Western Reserve University, USA}

\begin{abstract}

We report the detection of diffuse starlight in an extragalactic HI
cloud in the nearby Leo~I galaxy group. We detect the source, dubbed
BST1047+1156, in both broadband optical and GALEX ultraviolet light.
Spanning $\sim$ 2 kpc in radius, it has a peak surface brightness of
\mub=28.8 \magsec, making it the lowest surface brightness object ever
detected via integrated light. Although the object is extremely
gas-rich, with a gas fraction of $f_g=0.99$, its peak HI column density
is well below levels where star formation is typically observed in
galaxies. Nonetheless, BST1047+1156 shows evidence for young stellar
populations: along with the detected UV emission, the object is
extremely blue, with $B-V=0.14 \pm 0.09$. The object sports two tidal
tails and is found embedded within diffuse gas connecting the spiral
galaxy M96 to the group's extended HI Leo Ring. The nature of
BST1047+1156 is unclear. It could be a disrupting tidal dwarf, recently
spawned from star formation triggered in the Leo~I Group's tidal debris.
Alternatively, the object may have been a pre-existing galaxy --- the
most extreme example of a gas-rich field LSB known to date --- which had
a recent burst of star formation triggered by encounters in the group
environment.

\end{abstract}


\section{Introduction}

As optical imaging studies probe to deeper limits, the discovery of
galaxies at ever-lower low surface brightnesses continues to challenge
models of galaxy formation and evolution. Much attention has been paid
to the extremely diffuse galaxies found in cluster environments (\eg
Sandage \& Bingelli 1984; Impey \etal 1988; Caldwell 2006, van Dokkum
\etal 2015; Mihos \etal 2015), but in dense clusters it is difficult to
disentangle questions of galaxy formation from those of subsequent
dynamical evolution due to the various heating and stripping processes
at work in dense environments. In contrast, the gas-rich low surface
brightness (LSB) galaxies found in low density environments (\eg
McGaugh \& Bothun 1994; McGaugh \& de Blok 1997; Cannon \etal 2015,
Leisman \etal 2017) should have a less complicated evolutionary path,
and may better probe mechanisms of galaxy formation.

The high gas fractions found in field LSB galaxies (with $f_g \equiv
M_{gas}/(M_{gas}+M_*)$ as high as 0.98, Janowiecki \etal 2015; Ball
\etal 2018) argue that these systems are largely unevolved. Under
canonical models for galaxy formation, such galaxies may have
formed relatively late in the universe's history from the collapse of
high angular momentum material (\eg Dalcanton \etal 1997; Mo \etal 1998;
Amorisco \& Loeb 2016). Alternatively, some fraction of these systems
may have instead condensed within gas rich tidal debris  stripped from
galaxies during  close encounters (\eg Weilbacher \etal 2000; Duc \etal 2000; Bournaud
\etal 2004; Lelli \etal 2015). If these ``tidal dwarf'' galaxies survive
subsequent dynamical destruction, they would represent a population of
low mass, gas-rich galaxies distinct from classical dwarfs formed by
the collapse of material onto dark matter halos. 

Regardless of their formation history, such objects also probe star
formation processes in low density environments. Their low gas densities
are often at or below that required for wide-spread star formation (van
der Hulst \etal 1993; van Zee \etal 1997; Wyder \etal 2009), resulting
in sputtering and inefficient star formation histories (Schombert \etal
2001; Schombert \& McGaugh 2014, 2015). These effects may limit their
ability to form a substantial stellar disk, suggesting that absent
subsequent dynamical evolution there may well be a fundamental
``floor'' to galaxy surface brightness set by the inability to form stars
at low surface densities. Probing the lower limits to galaxy surface
brightness is thus critically important to models of both galaxy
formation and star formation.

Here we report the detection of diffuse starlight in an extragalactic
cloud in the nearby Leo~I galaxy group ($d=11$ Mpc; Graham \etal 1997;
Jang \& Lee 2017). We detect the system, named BST1047+1156 (hereafter
BST1047), in optical broadband and GALEX ultraviolet light and confirm
its kinematic association with tidal gas in the Leo I group. With an
extremely low peak surface brightness of \mub $\approx$ 28.8 \magsec,
BST1047 is the most diffuse object yet detected via integrated
starlight. The system may be the evolved counterpart of star forming
regions in the group's extended HI ring, a tidal dwarf possibly
undergoing disruption. Alternatively, it may be a pre-existing LSB
galaxy tidally perturbed by encounters within the group environment. In
this {\sl Letter}, we compare BST1047 to other types of LSB galaxies,
and consider various evolutionary scenarios to describe this extreme
system.

\section{Optical Detection and HI Mapping}

\begin{figure*}[]
\centerline{\includegraphics[width=7.0in]{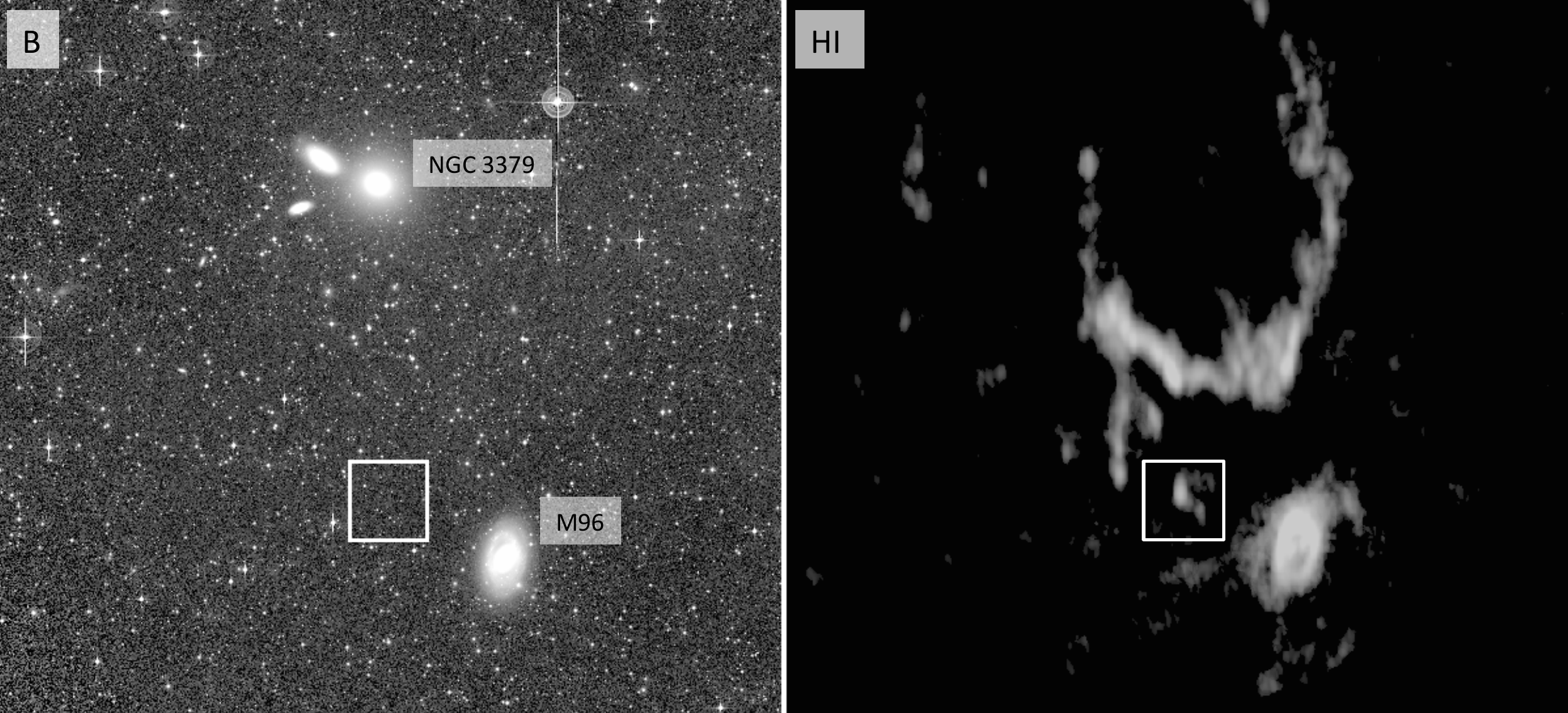}}
\caption{Wide field (1.5\degr$\times$1.6\degr) B-band image (left)
  and 21-cm HI map (right) of the Leo I group (from Watkins \etal 2014 and
  Oosterloo \etal 2010 respectively). BST1047 is located at the center of the
  white box, whose 9.5\arcmin$\times$9.5\arcmin\ size matches the field of 
  view of Figure~\ref{zoom}.
}
\label{imaging}
\end{figure*}

The optical imaging used here was originally taken as part of a deep
search for intragroup light in the Leo I Group, and is described in
detail in Watkins \etal (2014; hereafter W+14). We imaged Leo I over two
seasons with CWRU's Burrell Schmidt telescope, using a modified Johnson
B filter (Spring 2012) and Washington M filter (Spring 2013). We used a
dither-and-stack technique (W+14) with a total observing time of 13.7
and 10.25 hours in B and M, respectively. The resulting mosaics
(Figure~\ref{imaging}) covered $\sim 2.4\times2.4$ degree$^2$ at 1.45
arcsec pixel$^{-1}$ and have a liming depth of \mub=30.0 and \muv=29.5
\magsec\ after transformation to Johnson Vega magnitudes.

While W+14 discussed the properties of diffuse light on large scales in
the Leo I Group, BST1047 was only discovered when one of us (C.T.C.)
undertook a more detailed search for low surface brightness objects on
smaller scales. We detect BST1047 in both optical bands, projected
15.7\arcmin\ (50 kpc) ENE of the spiral galaxy M96. While the object is
extremely faint, with a peak surface brightness of $\mu_B = 28.8$, the
fact that it appears in {\sl independent} image stacks taken in
different seasons and comprised of many overlapping, dithered images
rules out the possibility of scattered light or instrumental artifacts.
By masking compact sources in the image and comparing the flux in the
object to that measured in 30\arcsec\ apertures scattered randomly
around the source, we find the object is detected at the 5.4$\sigma$ and
4.3$\sigma$ level in B and V, respectively. Figure~\ref{zoom} shows a
close-up view of BST1047 at different wavelengths, where the optical
images are shown both at full resolution and after being masked and
binned to enhance the diffuse light.

After the optical discovery, comparison of the imaging to HI mapping of
the Leo~I group (Schneider 1985, Schneider \etal 1986, 1989, Haan \etal
2009, Oosterloo \etal 2010) revealed an HI source spatially coincident with
the optical detection. Although the HI detection precedes our deep
optical study, we emphasize that our detection of BST1047 was a purely
optical one, with no prior knowledge by C.T.C. of the HI imaging data.
To our knowledge, this makes BST1047 the lowest surface brightness
object ever discovered via integrated light.

Figure~\ref{imaging} shows the Westerbork HI map of the Leo~I group
obtained after re-processing the observations of Oosterloo \etal (2010).
The spatial resolution is 105.0\arcsec $\times$39.2\arcsec\ (PA =
0\degr) and the noise level is 0.70 mJy beam$^{-1}$ for a velocity
resolution of 8.0 \kms. The HI data is shown in detail in
Figure~\ref{HIdata}. The HI line is only marginally resolved in
velocity, with an observed (Gaussian) line width of $\sigma = 9.0 \pm 1.0$ \kms.
While the object's north-south elongation is partly due to the
non-circular beam of the telescope, a small velocity gradient across the
main body indicates that the source is slightly resolved spatially as
well. On larger scales, the object is found near the gas seen extending
southward from the Leo Ring in Figure~\ref{imaging}, and is in fact
embedded in a stream of gas connecting the Ring to M96 at even lower
column densities (Schneider 1985; Schneider \etal 1989). BST1047 also
shows two diffuse HI tails stretching towards M96 to the west,
suggestive of tidal or ram pressure stripping. The object's systemic
velocity (970 \kms) is similar to that of M96 (897 \kms) and NGC~3379
(911 \kms), and unambiguously places it within the Leo~I group.

\section{Source Properties}

BST1047's extremely diffuse nature makes quantitative photometry
challenging. The object is roughly 35\arcsec\ in radius before dropping
below the limiting surface brightness of \mub=30.0 \magsec, and within this
radius are a number of compact sources. These sources span a range of
\bmv\ colors similar to star forming knots observed in other LSB
galaxies (Schombert \etal 2013), making it possible that some of these
sources could be intrinsic to the object itself. However, these sources
are spatially consistent with being background contaminants, as they are
not centrally concentated in the system but are found largely in the
outskirts at $r>25$\arcsec. {\sl Here we take the conservative approach
of masking these sources and measuring the photometric properties of the
diffuse light only.} Furthermore, given BST1047's irregular surface
brightness profile and low S/N, the object is not amenable to
traditional profile fitting. Instead, we give non-parametric estimates
of the limiting \mub=30 isophotal size ($R_{30}$), the effective radius
($R_e$) containing half the light encompassed within $R_{30}$, the mean
surface brightness within $R_e$ (\brackmueb), and the total integrated
magnitudes and colors. These values are reported in Table~\ref{props},
corrected for foreground reddening using A$_B$=0.09, E(B-V)=0.02
(Schlafly \& Finkbeiner 2011).

\begin{figure*}[]
\centerline{\includegraphics[width=7.0truein]{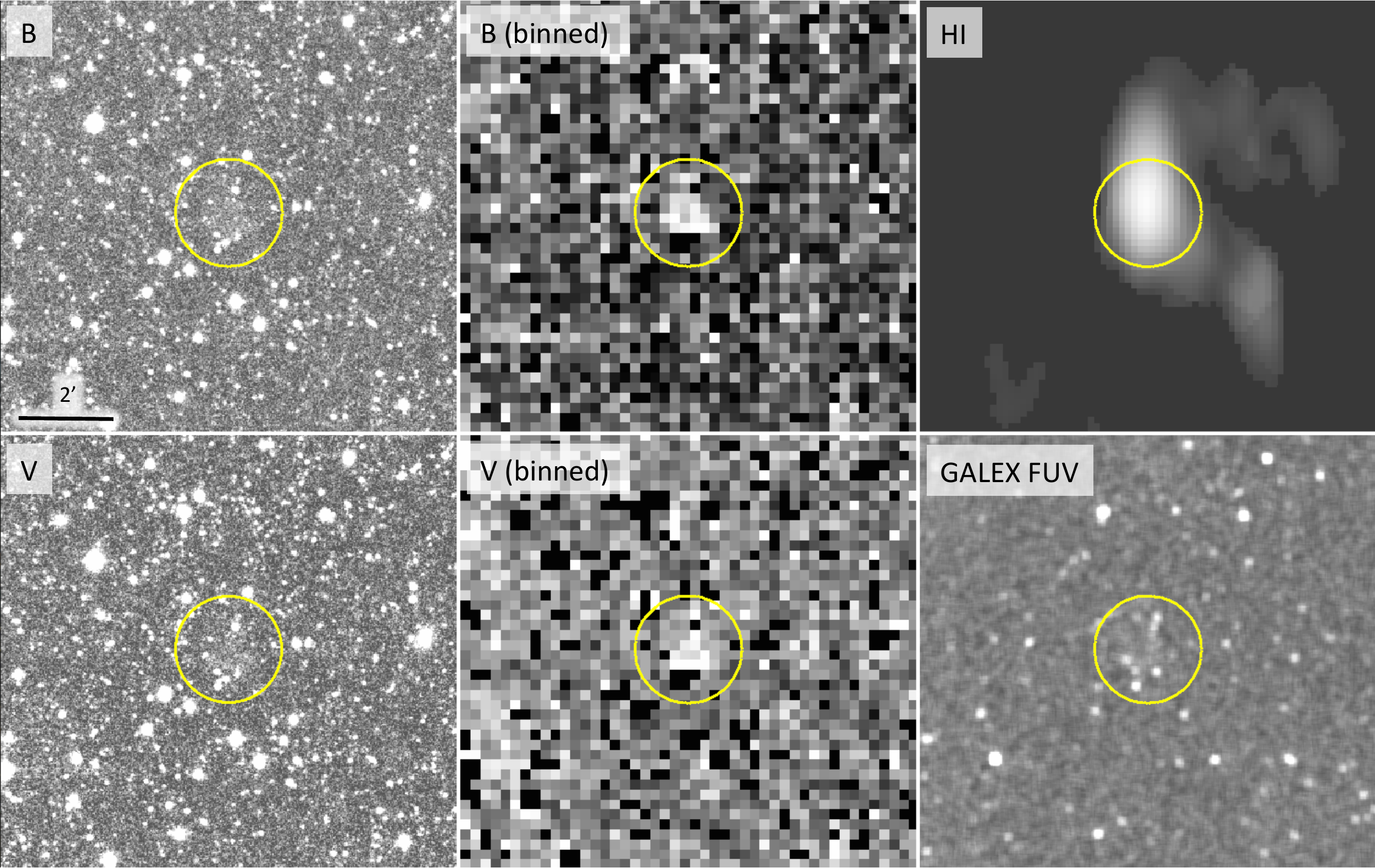}}
\caption{Imaging of BST1047 at multiple wavelengths. Optical imaging is
shown in the left panels; the center panels show the optical images
after masking compact sources and median binning to
13\arcsec$\times$13\arcsec\ resolution. The upper right panel shows the
HI map. The lower right panel shows the GALEX FUV image, boxcar smoothed
3$\times$3 (4.5\arcsec$\times$ 4.5\arcsec) to show faint detail. The
yellow circle has a radius of 70\arcsec, {\sl twice} the size of the
$R_{30}$ isophote. }
\label{zoom}
\end{figure*}

\begin{figure*}[]
\centerline{\includegraphics[width=7.0truein]{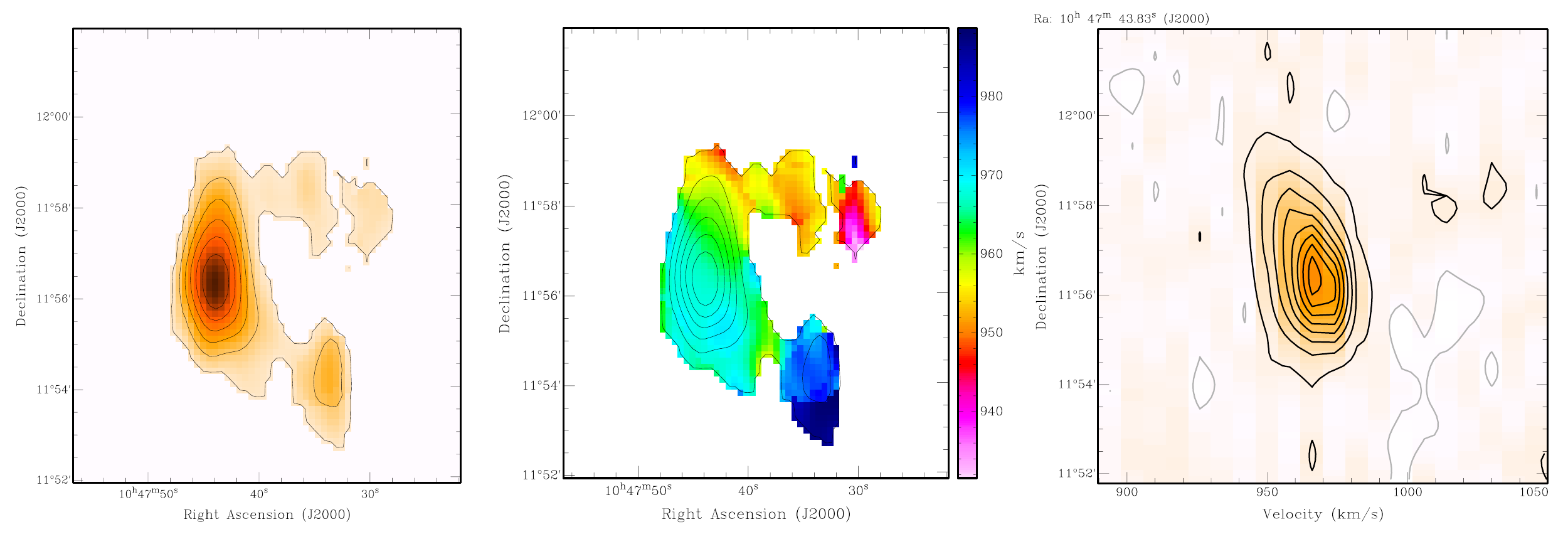}}
\caption{
Left: HI column density, with contours shown at 1.1, 3.3, 5.5, 7.7, 9.9,
and 11.1$\times 10^{19}$ cm$^{-2}$. Center: 2D velocity field. Right:
North/south position-velocity cut through the object.
}
\label{HIdata}
\end{figure*}

\begin{deluxetable}{lc}
\tabletypesize{\small}
\tablewidth{0pt}
\tablecaption{BST1047+1156 \label{props}}
\tablehead{\colhead{ } & \colhead{}}
\startdata
Optical center (J2000) & (10:47:43.8, +11:56:01) \\
R$_{30}$ & 35\arcsec\ (1.85 kpc) \\
R$_e$ & 22\arcsec\ (1.2 kpc) \\
\brackmueb & 28.9 mag arcsec$^{-2}$\\ 
$m_B, m_V$ (Vega) & $20.02^{+0.14}_{-0.13}$, $19.86^{+0.22}_{-0.16}$ \\
$(B-V)$ & $0.14^{+0.09}_{-0.09}$ \\
M$_B$  & $-10.2$  \\
L$_B$ & $1.85\times10^6$ \Lsun  \\
$m_{FUV}, m_{NUV}$ (AB) & $21.66^{+0.25}_{-0.17}$,  $21.19^{+0.20}_{-0.16}$\\
$(FUV-NUV)$ &  $0.47^{+0.29}_{-0.26}$ \\
HI peak (J2000) & (10:47:43.8, +11:56:18) \\
HI flux &  1.59 Jy km/s \\
HI mass & $4.5\times 10^7$ \Msun \\
$\Sigma_{HI,peak}$ & $1.4\times 10^{20}$ cm$^{-2}$ \\
$v_{sys}$ & 970 \kms \\
\enddata
\label{props}
\end{deluxetable}

\begin{figure}[]
\centerline{\includegraphics[width=3.5truein]{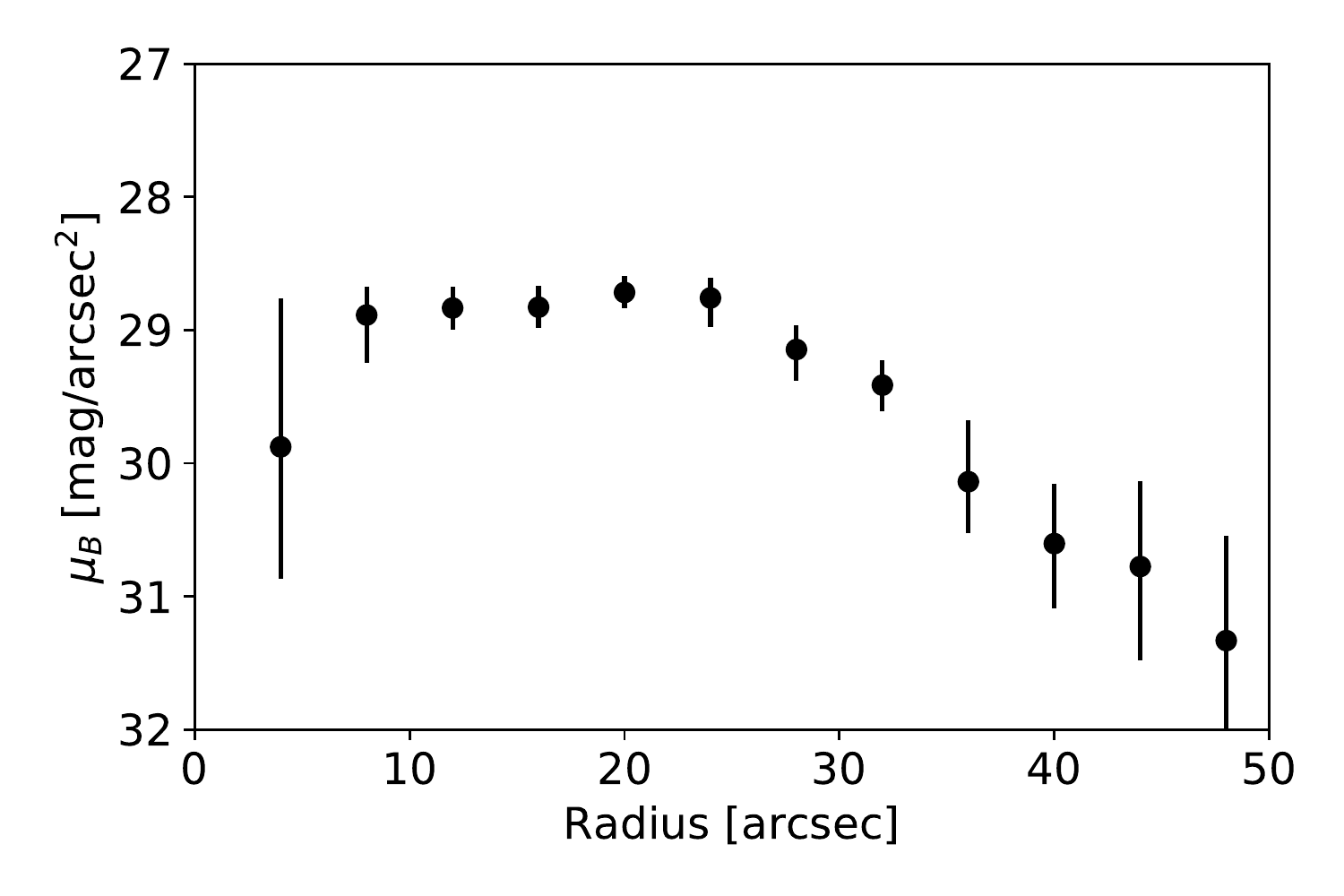}}
\caption{
B-band surface brightness profile of BST1047.
}
\label{surfbprof}
\end{figure}

Figure~\ref{surfbprof} shows BST1047's surface brightness profile, with
errorbars reflecting both the $\sim$ 5\arcsec\ positional uncertainty in 
the optical center as well as background uncertainties (measured from
30\arcsec\ sky apertures scattered randomly in the surrounding field).
Not only is the object diffuse, with an average surface brightness of
\brackmueb = 28.9 \magsec, it has a flat luminosity profile with no
evidence for central concentration. Summing the diffuse light within
$R_{30}$ yields a total magnitude of $m_B=20.02^{+0.14}_{-0.13}$, and an
extremely blue color of $(B-V)=0.14^{+0.09}_{-0.09}$.

Given the object's very blue color, we also searched for it in GALEX
ultraviolet imaging (Figure~\ref{zoom}). In both FUV and NUV we recover
many of the compact sources, along with weak, diffuse UV
flux detected at the 3.4$\sigma$ (FUV) and 3.7$\sigma$ (NUV) levels.
This diffuse UV light has a $FUV-NUV$ color of $0.47^{+0.29}_{-0.26}$,
after correction for foreground reddening (Thilker \etal 2009). The
detected FUV emission suggests the system has experienced recent
star formation over the past few hundred Myr (\ie the FUV emitting
lifetime of young stellar populations; Kennicutt \& Evans 2012).

BST1047 has a total HI flux of 1.59 Jy \kms, giving an HI mass of
M$_{\rm HI}=4.5\times10^7$ \Msun\llap, or a total gas mass (correcting
for helium) of M$_{gas}=1.33\times{\rm M_{HI}}=6\times10^7$ \Msun\llap.
Adopting a stellar mass-to-light ratio of $M_*/L_B = 0.2$, appropriate
for stellar populations with the colors measured here (Bell \& de Jong
2001), yields a stellar mass of $M_* = 3.7\times10^5$ \Msun and gas
fraction $f_g=0.99$, rivaling the most extreme measured for any galaxy
(\eg Janowiecki \etal 2015). However, the gas is quite
diffuse, with peak HI column density $1.4\times10^{20}$ cm$^{-2}$, well
below levels at which star formation typically occurs (\eg Bigiel \etal
2008, 2010; Krumholz \etal 2009; Clark \& Glover 2014). This result is
seemingly at odds with the blue optical colors and FUV emission, both
indicators of recent star formation. Given the relatively large beam
size of the Westerbork data, pockets of higher density gas could exist
on smaller scales to drive star formation, although no H$\alpha$
emission was detected in deep narrowband imaging of the group by Donahue
\etal (1995).

Indeed, the properties of BST1047 are so extreme that they may call into
question the likelihood that we are seeing bona-fide starlight at all.
At these faint surface brightnesses, deep imaging surveys are plagued by
contamination due to scattered light from Milky Way dust (the Galactic
cirrus; see, \eg Mihos \etal 2017), but this is not a concern here. The
object is not detected in infrared emission (a common tracer of Galactic
dust) in either the WISE 12$\mu$m or IRIS 100$\mu$m maps (Meisner \&
Finkbeiner 2014; Miville-Desch\^enes \& Lagache 2005, respectively), and
the measured HI velocity of the system lies well outside the range of
Milky Way high velocity clouds (Wakker \& van Woerden 1997). However, a
second possibility is scattering from dust in the object at the Leo~I
distance. A simple calculation shows this to be unlikely. The closest
source of optical photons is M96, projected 50 kpc to the west. If we
take the flux impingent on BST1047 from M96 and use the dust scattering
model of Draine (2003) to scatter that light along our line of sight,
the {\sl maximum} observed surface brightness achieved is $\mu_B \sim
31$ \magsec, significantly fainter than observed. Since dust
preferentially forward-scatters light, this maximum surface brightness
is achieved only in the specific situation where BST1047 lies 50 kpc in
front of M96 (as well as 50 kpc in transverse projection); other
geometries predict even lower surface brightnesses due either to the
scattering phase function or a larger separation from M96 reducing the
impingent stellar light on BST1047. Furthermore, the Draine model adopts
a Milky Way dust-to-gas ratio and likely overestimates the dust content
of BST1047 by a significant amount. As a result, scattering of M96
starlight from dust within BST1047 is unlikely to explain the diffuse
light we observe.

\section{Discussion}

What is BST1047? One natural comparison is to the star forming knots
embedded in the Leo Ring to the north (Thilker \etal 2009; Michel-Dansac
\etal 2010; W+14). Compared to these complexes, BST1047 is similar in
size and HI mass, but fainter in FUV light and somewhat redder in UV
color ($(FUV-NUV)_{\rm knots} = -0.3$ to +0.05). This suggests the
system may be an older, more evolved version of the star forming
complexes seen in the Leo Ring, observed after star formation has
largely died out. Using the FUV star formation rate (SFR) estimator of
Hunter \etal (2010) gives a total SFR for BST1047 of $\sim
1.4\times10^{-4}$ \Msun yr$^{-1}$ (lower by a factor of 4--5 compared to
the star forming knots in the Ring). At this rate, the time to build the
observed stellar mass is $\sim 3$ Gyr, but continuous star formation
would produce stellar populations too red to match the observed optical
color. A rapidly declining SFR is also indicated by the lack of detected
H$\alpha$ in the imaging of Donahue \etal (1995) -- adopting the
H$\alpha$-SFR calibration of Murphy \etal (2011) gives an upper limit to
the {\sl current} SFR of $\sim 5\times10^{-5}$ \Msun\ yr$^{-1}$, down by
nearly a factor of three compared to that inferred from the FUV light,
which traces the SFR over longer 100--200 Myr timescales. Hence the FUV
emission and blue colors seen in BST1047 could be tracing a {\sl
post-burst} stellar population formed by recent star formation. A
comparison of the optical and UV colors to stellar population synthesis
models (Leitherer \etal 1999; Bruzual \& Charlot 2003) shows consistency
with population ages of 200--600 Myr.

The similarity in HI mass and size between Leo Ring clumps and known
dwarf galaxies has led several authors to suggest the Ring clumps may be
tidal dwarf galaxies in formation (\eg Schneider \etal 1986; Thilker
\etal 2009). BST1047 may be a similar and somewhat older object
transitioning to a quiescent evolutionary stage. Under this scenario,
the BST1047's spatial and kinematic association with the diffuse gas
connecting the Leo Ring to M96 suggest it may have formed during an
interaction between M96 and the Ring. The stellar population age
inferred above sets the timescale for such an event at $\sim$ 200--600
Myr ago, somewhat shorter than the $\sim$ 1 Gyr old interaction age
hypothesized by Michel-Dansac \etal (2010) in their model of the
formation of the Leo Ring. If that model is correct, the interaction
that gave rise to BST1047 may be secondary to that which formed the Ring
itself.

Tidal dwarfs are expected to be free of dark matter, having
condensed due to the compression of tidal gas. If the
kinematics of BST1047 trace the gravitational potential, we can
infer its dynamical mass using the estimator of Hoffman \etal (1999): 
$$M_{dyn} =2.325\times10^5 
\Big({{V_{rot}^2 +3\sigma^2}\over{\rm km^2\ s^{-2}}}\Big)
\Big({{R}\over{\rm kpc}}\Big) M_\odot.$$
For BST1047, several factors complicate this analysis. First, it is
unclear if the observed velocity gradient reflects tidal shear or true
rotation, and, furthermore, if the system is rotating we have little
information on its inclination. Also, the HI line width is only slightly
resolved over instrumental, making a reliable measurement of the
intrinsic velocity dispersion difficult. Nonetheless, if we presume the
gradient is rotational, we can estimate $V_{rot}$ as half the
north-south velocity difference measured at 20\% peak HI intensity (the
second positive contour in Figure~\ref{HIdata}c), which gives $V_{rot}$
= 8 \kms. Taking the quadrature difference between the observed Gaussian
line width and the instrumental dispersion yields an intrinsic velocity
dispersion of $\sigma=4$ \kms. Using these values along with the
$R_{30}=1.85$ kpc isophotal radius yields $M_{dyn}=4.8\times10^7$
\Msun\llap, comparable to the observed baryonic mass ($6\times10^7$
\Msun\llap). Although consistent with the tidal dwarf scenario in which
baryons provide all the dynamical mass, our mass estimate is very
uncertain; if the system is observed largely face-on, any inclination
correction would increase $M_{dyn}$, while a lack of rotation would drop
$M_{dyn}$ to only one-third of the baryonic mass.

The survival of a tidal dwarf depends on a complicated balance between
its binding mass (set only by the baryons), energy injection from
massive stars, and dynamical stripping by the local environment.
BST1047's HI tidal tails indicate the object is being stripped as it
moves through the Leo~I group. This stripping, coupled possibly with
energy injection from the initial starburst, could have driven gas
densities down to the current levels that inhibit continuing star
formation. The initial compression of gas that formed BST1047 may now be
``rebounding," and the stars --- following the gravitational potential
dominated by the gas --- would disperse as well, leading to the flat
surface brightness profile seen in Figure~\ref{surfbprof}. Thus that the
object we see today might only be transitory --- recently formed, but in
the process of disruption in the group environment.

In this scenario, then, BST1047 might be a {\sl failed} (or failing)
tidal dwarf, with little or no ongoing star formation and insufficient
binding mass to ensure its survival. While prospective tidal dwarfs
in some other systems appear to have sufficient mass to survive subsequent
destruction (\eg Lelli \etal 2015), BST1047 suggests that not all
systems might be so lucky. Gas-rich interactions could lead to the
formation and expulsion of fragile dwarfs from their host galaxies,
after which they are destroyed and dispersed into intergalactic space,
adding to the population of intragroup or intracluster stars found in
many high density environments. In the Leo~I group we may be witnessing
the full evolutionary path of these objects, from formation (the
star-forming knots) through disruption (BST1047) and ending
with dispersal in the form of the extended and faint ``orphan'' stellar
streams observed in group (see Figure~2 of W+14).

However, an alternative explanation might be that rather than being a tidally-spawned
object, BST1047 is instead a pre-existing, extremely diffuse LSB galaxy,
perhaps a satellite companion of M96, that has been caught in a tidal
interaction within the group. If the object is an M96 satellite,
another analogue might be the diffuse star-forming
satellites in and around the Local Group, such as Leo T (Irwin \etal
2007; Ryan-Weber \etal 2008) and Leo P (McQuinn \etal 2013, 2015).
Compared to these systems, BST1047 is significantly larger (\re $\sim$ 1
kpc, compared to 0.1--0.2 kpc for Leo T and Leo P) and more luminous (by
1--2 magnitudes), with a much higher gas mass (by two orders of
magnitude). However, the key factor that differentiates BST1047 is not
just size but density. Both Leo P and Leo T have central surface
brightnesses of \muv $\approx$ 24.5 \magsec, 50 times higher in
luminosity density than BST1047. And while Leo P and Leo T are
significantly lower in gas mass, their compact nature means that their
HI column densities are actually several times {\sl higher} than those
found in BST1047. In BST1047, the low gas density likely means the
object has been even less efficient than Leo P or Leo T at forming
stars, and may explain its extremely high gas fraction ($f_g=0.99$,
compared to $f_g = 0.8$ and 0.6 for Leo T and Leo P, respectively;
Ryan-Weber \etal 2008; McQuinn \etal 2015).

Given BST1047's dissimilarity to the faint, diffuse star-forming
satellites of the Local Group, another comparison would be to larger low
surface brightness galaxies. BST1047's extremely low surface brightness
is similar to the UDGs known to populate group and cluster environments,
but these UDGs are typically red with little evidence for recent star
formation (van Dokkum \etal 2015; Koda \etal 2015; Lee \etal 2017),
making them a poor match to BST1047. Instead, blue gas-rich field LSBs
may make a better comparison sample. BST1047 shares many of the
properties of field LSBs, but taken to extremes. For example, the most
diffuse LSB in the ``(Almost) Dark'' sample is Coma~P, with peak surface
brightness $\mu_{B} \approx 26.6$ \magsec, \bmv=0.13, and gas fraction
$f_g=0.97$ (Janowiecki \etal 2015; Ball \etal 2018). BST1047 is similar
in optical size and color to Coma~P, but an order of magnitude lower in
luminosity density. And while the total gas mass of both systems is
comparable, BST1047's peak HI column of $\sim$ 1 \Msun pc$^{-2}$ makes
it lower by a factor of four compared to Coma~P.

Field LSBs are also marked by stochastic and inefficient star formation
histories (Schombert \& McGaugh 2014, 2015),
likely due to their low gas densities, often below that thought
necessary to trigger widespread star formation (\eg van der Hulst \etal
1993; McGaugh \& de Blok 1997; Wyder \etal 2009). Such a SFR must also
be true for BST1047, given the need for a rapidly declining SFR to
explain the optical/UV colors, stellar mass, and FUV and H$\alpha$ star
formation tracers discussed previously. While the gas densities in
BST1047 are likely too low to sustain star formation, a recent,
short-lived, and now-declining burst of star formation could have been
triggered by a tidal encounter in the Leo~I group. BST1047's association
with the Leo Ring's southern HI spur, along with the object's own HI
tidal tails, are consistent with such a scenario, possibly involving the
nearby spiral M96.

In this scenario, then, BST1047 would simply represent the most diffuse
tail of the field LSB galaxy population, involved in an interaction in
the group environment. If so, the object should follow the
well-established baryonic Tully-Fisher (BTF) relationship (McGaugh \etal
2000; McGaugh 2011). Given its baryonic mass of $6\times10^7$ \Msun, the
BTF of McGaugh (2011) predicts a velocity width of of $\Delta v \approx
2\times v_c = 67$ \kms. This is significantly larger than the observed
16 \kms\ HI velocity gradient, but could be accommodated if the galaxy
is observed largely face on, with an inclination of roughly 15\degr.
Unfortunately, while the imaging in Figure~\ref{zoom} does not suggest a
highly inclined geometry, BST1047's extremely diffuse nature and
irregular morphology makes inclination estimates difficult.

Discriminating between these various scenarios to explain BST1047 will require additional
data. Of particular use would be detailed 21-cm mapping of the system,
to resolve whether the velocity gradient is a sign of rotational motion
or tidal shear. Similarly, deep imaging of BST1047 by {\sl HST} or {\sl
JWST} would yield information on the age and metallicity of BST1047's
stellar populations --- detecting an old red giant branch would provide
strong argument against the tidal dwarf scenario. With such extreme
properties compared to known extragalactic objects, BST1047 will provide
important constraints on processes driving the formation, evolution, and
destruction of diffuse galaxies in group environments.

\acknowledgments

We thank Stacy McGaugh and the referee for many helpful suggestions. This work was supported
in part through NSF grant 1108964 to JCM. The Westerbork Synthesis Radio
Telescope is operated by the ASTRON (Netherlands Institute for Radio
Astronomy) with support from the Netherlands Foundation for Scientific
Research (NWO).

\facility{CWRU:Schmidt, WSRT}


\begin{thebibliography}{}



\bibitem[Amorisco \& Loeb(2016)]{2016MNRAS.459L..51A} Amorisco, N.~C., \& Loeb, A.\ 2016, \mnras, 459, L51 


\bibitem[Ball et al.(2018)]{2018AJ....155...65B} Ball, C., Cannon, J.~M., Leisman, L., et al.\ 2018, \aj, 155, 65 


\bibitem[Bell \& de Jong(2001)]{2001ApJ...550..212B} Bell, E.~F., \& de Jong, R.~S.\ 2001, \apj, 550, 212 


\bibitem[Bigiel et al.(2010)]{2010AJ....140.1194B} Bigiel, F., Leroy, A., Walter, F., et al.\ 2010, \aj, 140, 1194 


\bibitem[Bigiel et al.(2008)]{2008AJ....136.2846B} Bigiel, F., Leroy, A., Walter, F., et al.\ 2008, \aj, 136, 2846 


\bibitem[Bournaud et al.(2004)]{2004A&A...425..813B} Bournaud, F., Duc, P.-A., Amram, P., Combes, F., \& Gach, J.-L.\ 2004, \aap, 425, 813 


\bibitem[Bruzual \& Charlot(2003)]{2003MNRAS.344.1000B} Bruzual, G., \& Charlot, S.\ 2003, \mnras, 344, 1000 


\bibitem[Caldwell(2006)]{2006ApJ...651..822C} Caldwell, N.\ 2006, \apj, 651, 822 


\bibitem[Cannon et al.(2015)]{2015AJ....149...72C} Cannon, J.~M., Martinkus, C.~P., Leisman, L., et al.\ 2015, \aj, 149, 72 


\bibitem[Clark \& Glover(2014)]{2014MNRAS.444.2396C} Clark, P.~C., \& Glover, S.~C.~O.\ 2014, \mnras, 444, 2396 


\bibitem[Dalcanton et al.(1997)]{1997ApJ...482..659D} Dalcanton, J.~J., Spergel, D.~N., \& Summers, F.~J.\ 1997, \apj, 482, 659 


\bibitem[Donahue et al.(1995)]{1995ApJ...450L..45D} Donahue, M., Aldering, G., \& Stocke, J.~T.\ 1995, \apjl, 450, L45 


\bibitem[Draine(2003)]{2003ApJ...598.1017D} Draine, B.~T.\ 2003, \apj, 598, 1017 


\bibitem[Duc et al.(2000)]{2000AJ....120.1238D} Duc, P.-A., Brinks, E., Springel, V., et al.\ 2000, \aj, 120, 1238 


\bibitem[Graham et al.(1997)]{1997ApJ...477..535G} Graham, J.~A., Phelps, R.~L., Freedman, W.~L., et al.\ 1997, \apj, 477, 535 


\bibitem[Haan et al.(2009)]{2009ApJ...692.1623H} Haan, S., Schinnerer, E., Emsellem, E., et al.\ 2009, \apj, 692, 1623 


\bibitem[Hoffman et al.(1999)]{1999AJ....117..811H} Hoffman, G.~L., Lu, N.~Y., Salpeter, E.~E., \& Connell, B.~M.\ 1999, \aj, 117, 811 


\bibitem[Hunter et al.(2010)]{2010AJ....139..447H} Hunter, D.~A., Elmegreen, B.~G., \& Ludka, B.~C.\ 2010, \aj, 139, 447 


\bibitem[Impey et al.(1988)]{1988ApJ...330..634I} Impey, C., Bothun, G., \& Malin, D.\ 1988, \apj, 330, 634 


\bibitem[Irwin et al.(2007)]{2007ApJ...656L..13I} Irwin, M.~J., Belokurov, V., Evans, N.~W., et al.\ 2007, \apjl, 656, L13 


\bibitem[Jang \& Lee(2017)]{2017ApJ...836...74J} Jang, I.~S., \& Lee, M.~G.\ 2017, \apj, 836, 74 


\bibitem[Janowiecki et al.(2015)]{2015ApJ...801...96J} Janowiecki, S., Leisman, L., J{\'o}zsa, G., et al.\ 2015, \apj, 801, 96 


\bibitem[Kennicutt \& Evans(2012)]{2012ARA&A..50..531K} Kennicutt, R.~C., \& Evans, N.~J.\ 2012, \araa, 50, 531 


\bibitem[Koda et al.(2015)]{2015ApJ...807L...2K} Koda, J., Yagi, M., Yamanoi, H., \& Komiyama, Y.\ 2015, \apjl, 807, L2 


\bibitem[Krumholz et al.(2009)]{2009ApJ...699..850K} Krumholz, M.~R., McKee, C.~F., \& Tumlinson, J.\ 2009, \apj, 699, 850 


\bibitem[Lee et al.(2017)]{2017ApJ...844..157L} Lee, M.~G., Kang, J., Lee, J.~H., \& Jang, I.~S.\ 2017, \apj, 844, 157 


\bibitem[Leisman et al.(2017)]{2017ApJ...842..133L} Leisman, L., Haynes, M.~P., Janowiecki, S., et al.\ 2017, \apj, 842, 133 


\bibitem[Leitherer et al.(1999)]{1999ApJS..123....3L} Leitherer, C., Schaerer, D., Goldader, J.~D., et al.\ 1999, \apjs, 123, 3 


\bibitem[Lelli et al.(2015)]{2015A&A...584A.113L} Lelli, F., Duc, P.-A., Brinks, E., et al.\ 2015, \aap, 584, A113 


\bibitem[McGaugh \& Bothun(1994)]{1994AJ....107..530M} McGaugh, S.~S., \& Bothun, G.~D.\ 1994, \aj, 107, 530 


\bibitem[McGaugh et al.(2000)]{2000ApJ...533L..99M} McGaugh, S.~S., Schombert, J.~M., Bothun, G.~D., \& de Blok, W.~J.~G.\ 2000, \apjl, 533, L99 


\bibitem[McGaugh(2011)]{2011PhRvL.106l1303M} McGaugh, S.~S.\ 2011, Physical Review Letters, 106, 121303 


\bibitem[McGaugh \& de Blok(1997)]{1997ApJ...481..689M} McGaugh, S.~S., \& de Blok, W.~J.~G.\ 1997, \apj, 481, 689 


\bibitem[McQuinn et al.(2013)]{2013AJ....146..145M} McQuinn, K.~B.~W., Skillman, E.~D., Berg, D., et al.\ 2013, \aj, 146, 145 


\bibitem[McQuinn et al.(2015)]{2015ApJ...812..158M} McQuinn, K.~B.~W., Skillman, E.~D., Dolphin, A., et al.\ 2015, \apj, 812, 158 


\bibitem[Meisner \& Finkbeiner(2014)]{2014ApJ...781....5M} Meisner, A.~M., \& Finkbeiner, D.~P.\ 2014, \apj, 781, 5 


\bibitem[Michel-Dansac et al.(2010)]{2010ApJ...717L.143M} Michel-Dansac, L., Duc, P.-A., Bournaud, F., et al.\ 2010, \apjl, 717, L143 


\bibitem[Mihos et al.(2015)]{2015ApJ...809L..21M} Mihos, J.~C., Durrell, P.~R., Ferrarese, L., et al.\ 2015, \apjl, 809, L21 


\bibitem[Mihos et al.(2017)]{2017ApJ...834...16M} Mihos, J.~C., Harding, P., Feldmeier, J.~J., et al.\ 2017, \apj, 834, 16 


\bibitem[Miville-Desch{\^e}nes \& Lagache(2005)]{2005ApJS..157..302M} Miville-Desch{\^e}nes, M.-A., \& Lagache, G.\ 2005, \apjs, 157, 302 


\bibitem[Mo et al.(1998)]{1998MNRAS.295..319M} Mo, H.~J., Mao, S., \& White, S.~D.~M.\ 1998, \mnras, 295, 319 


\bibitem[Murphy et al.(2011)]{2011ApJ...737...67M} Murphy, E.~J., Condon, J.~J., Schinnerer, E., et al.\ 2011, \apj, 737, 67 


\bibitem[Oosterloo et al.(2010)]{2010MNRAS.409..500O} Oosterloo, T., Morganti, R., Crocker, A., et al.\ 2010, \mnras, 409, 500 


\bibitem[Ryan-Weber et al.(2008)]{2008MNRAS.384..535R} Ryan-Weber, E.~V., Begum, A., Oosterloo, T., et al.\ 2008, \mnras, 384, 535 


\bibitem[Sandage \& Binggeli(1984)]{1984AJ.....89..919S} Sandage, A., \& Binggeli, B.\ 1984, \aj, 89, 919 


\bibitem[Schlafly \& Finkbeiner(2011)]{2011ApJ...737..103S} Schlafly, E.~F., \& Finkbeiner, D.~P.\ 2011, \apj, 737, 103 


\bibitem[Schneider(1985)]{1985ApJ...288L..33S} Schneider, S.\ 1985, \apjl, 288, L33 


\bibitem[Schneider et al.(1986)]{1986AJ.....91...13S} Schneider, S.~E., Salpeter, E.~E., \& Terzian, Y.\ 1986, \aj, 91, 13 


\bibitem[Schneider et al.(1989)]{1989AJ.....97..666S} Schneider, S.~E., Skrutskie, M.~F., Hacking, P.~B., et al.\ 1989, \aj, 97, 666 


\bibitem[Schombert \& McGaugh(2015)]{2015AJ....150...72S} Schombert, J., \& McGaugh, S.\ 2015, \aj, 150, 72 


\bibitem[Schombert \& McGaugh(2014)]{2014PASA...31...36S} Schombert, J., \& McGaugh, S.\ 2014, \pasa, 31, e036 


\bibitem[Schombert et al.(2013)]{2013AJ....146...41S} Schombert, J., McGaugh, S., \& Maciel, T.\ 2013, \aj, 146, 41 


\bibitem[Schombert et al.(2001)]{2001AJ....121.2420S} Schombert, J.~M., McGaugh, S.~S., \& Eder, J.~A.\ 2001, \aj, 121, 2420 


\bibitem[Thilker et al.(2009)]{2009Natur.457..990T} Thilker, D.~A., Donovan, J., Schiminovich, D., et al.\ 2009, \nat, 457, 990 


\bibitem[van der Hulst et al.(1993)]{1993AJ....106..548V} van der Hulst, J.~M., Skillman, E.~D., Smith, T.~R., et al.\ 1993, \aj, 106, 548 


\bibitem[van Dokkum et al.(2015)]{2015ApJ...798L..45V} van Dokkum, P.~G., Abraham, R., Merritt, A., et al.\ 2015, \apjl, 798, L45 


\bibitem[van Zee et al.(1997)]{1997AJ....113.1618V} van Zee, L., Haynes, M.~P., Salzer, J.~J., \& Broeils, A.~H.\ 1997, \aj, 113, 1618 


\bibitem[Wakker \& van Woerden(1997)]{1997ARA&A..35..217W} Wakker, B.~P., \& van Woerden, H.\ 1997, \araa, 35, 217 


\bibitem[Watkins et al.(2014)]{2014ApJ...791...38W} Watkins, A.~E., Mihos, J.~C., Harding, P., \& Feldmeier, J.~J.\ 2014, \apj, 791, 38 


\bibitem[Weilbacher et al.(2000)]{2000A&A...358..819W} Weilbacher, P.~M., Duc, P.-A., Fritze v.~Alvensleben, U., Martin, P., \& Fricke, K.~J.\ 2000, \aap, 358, 819 


\bibitem[Wyder et al.(2009)]{2009ApJ...696.1834W} Wyder, T.~K., Martin, D.~C., Barlow, T.~A., et al.\ 2009, \apj, 696, 1834 

\end{thebibliography}
\end{document}